\documentclass[12pt,a4paper,fleqn]{article}
\usepackage[T1]{fontenc}
\usepackage{amsmath}
\usepackage{amssymb} 
\usepackage{bbm} 
\usepackage[small]{caption2} 
\usepackage{graphicx} 
\usepackage{mathrsfs}	
\usepackage[small,loose]{subfigure}  
\usepackage{cite} 
\advance \headheight by 3.0truept       

\DeclareMathOperator{\im}{Im}
\DeclareMathOperator{\tr}{tr}
\DeclareMathOperator{\diag}{diag}

\newcommand{\CenterEps}[2][1]{\ensuremath{\vcenter{\hbox{\includegraphics[scale=#1]{#2.eps}}}}} 
\newcommand{\D}{\mathrm{d}}
\newcommand{\I}{\mathrm{i}}

\newcommand{\SO}[1]{\ensuremath{\mathrm{SO}(#1)}}
\newcommand{\SU}[1]{\ensuremath{\mathrm{SU}(#1)}}
\newcommand{\U}[1]{\ensuremath{\mathrm{U}(#1)}}
\newcommand{\Z}[1]{\ensuremath{\mathbbm{Z}_{#1}}} 

\allowdisplaybreaks

\begin{document}
\thispagestyle{empty}
\begin{titlepage}

\begin{flushright}
TUM-HEP 722/09
\end{flushright}

\vspace*{1.0cm}

\begin{center}
\Huge\textbf{Gauge--top unification}
\end{center}
\vspace{1cm}
 \center{
\textbf{
Pierre Hosteins\footnote[1]{Email: \texttt{pierre.hosteins@ph.tum.de}}, 
Rolf Kappl\footnote[2]{Email: \texttt{rolf.kappl@ph.tum.de}}, 
Michael Ratz\footnote[3]{Email: \texttt{mratz@ph.tum.de}}, 
Kai Schmidt-Hoberg\footnote[4]{Email: \texttt{kschmidt@ph.tum.de}}
}
}
\\[5mm]
\begin{center}
\textit{\small
Physik-Department T30, Technische Universit\"at M\"unchen, \\
James-Franck-Stra\ss e, 85748 Garching, Germany
}
\end{center}

\date{20.05.2009}
\vspace{1cm}

\begin{abstract}
Higher-dimensional models of grand unification allow us to relate  the top
Yukawa coupling $y_t$ to the gauge coupling $g$. The tree level relation $y_t=g$
at the scale of grand unification implies, in the framework of the MSSM, a
rather  small ratio of Higgs expectation values $\tan\beta$.  We find that, in
the presence of localized Fayet-Iliopoulos terms,  $y_t$ is suppressed against
$g$ because the bulk fields acquire non-trivial  profiles whose overlap is
smaller than in the case of flat profiles. This increases the prediction for
$\tan\beta$ to moderately large values. Thus $\tan\beta$ is related to the
geometry of compact space. We also discuss explicit realizations of such
settings in orbifold compactifications of the heterotic string. It turns out
that anisotropic compactifications, allowing for an orbifold GUT interpretation,
are favored.
\end{abstract}

\end{titlepage}

\newpage

\section{Introduction}

The coupling strengths governing the interactions of the standard model (SM)
exhibit a very peculiar pattern: on the one hand, the gauge and top Yukawa
couplings are of order one, on the other hand all other Yukawa couplings are
suppressed. This might tell us that couplings come in two classes with
fundamentally different origin. In this study we shall investigate the scenario
of ``gauge-top unification'' in which the top Yukawa coupling $y_t$ and the
gauge couplings $g_a$ ($1\le a\le3$) unify at a high scale. 
Our analysis will be based on the minimal supersymmetric extension of the
standard model (MSSM), as it appears to provide us with the most compelling
scenario of gauge coupling unification, and therefore fits very nicely to the
concept of grand unified theories (GUTs) \cite{Georgi:1974sy,Fritzsch:1974nn}.

Arguably, the most compelling realizations of GUTs incorporate extra dimensions.
In fact, string-theoretic
orbifolds~\cite{Dixon:1985jw,Dixon:1986jc,Ibanez:1986tp,Ibanez:1987sn,Casas:1987us,Casas:1988hb,Font:1988tp,Font:1989aj}
and field-theoretic orbifold
GUTs~\cite{Kawamura:1999nj,Kawamura:2000ev,Altarelli:2001qj,%
Hall:2001pg,Hebecker:2001wq,Asaka:2001eh,Hall:2001xr,Burdman:2002se} allow us to
retain the beautiful aspects of grand unification while avoiding the notorious
problems. Merging both
approaches~\cite{Kobayashi:2004ud,Forste:2004ie,Kobayashi:2004ya,Buchmuller:2004hv}
has lead us to explicit string-derived models which reproduce the MSSM in their
low-energy limit and have a straightforward orbifold GUT
interpretation~\cite{Buchmuller:2005jr,Buchmuller:2006ik,Lebedev:2006kn,Buchmuller:2007qf,Lebedev:2007hv,Buchmuller:2008uq}
(for reviews see \cite{Ratz:2007my,Nilles:2008gq}).

In this paper we shall study scenarios in which $y_t$ arises from gauge
interactions in more than four dimensions. We will discuss settings in which the
Higgs fields arise from extra components of the gauge multiplet, enforcing the
tree-level relation
\begin{equation}\label{eq:yteqg}
 y_t~=~g
\end{equation} 
at the compactification scale~\cite{Burdman:2002se,Gogoladze:2003ci}. As we
shall see, relation \eqref{eq:yteqg} implies, together with the updated
mass of the top quark \cite{:2009ec}, uncomfortably small values for the ratio
of Higgs expectation values $\tan\beta$. However, following earlier work by Lee,
Nilles and Zucker \cite{Lee:2003mc} we find that localized Fayet-Iliopoulos
terms, which are generically present in these compactifications, always reduce
the value of $y_t$ at the compactification scale, thus increasing the prediction
for $\tan\beta$ to moderately large (or even large) values. This is
because the bulk hypermultiplets attain non-trivial profiles whose overlap is
always smaller than in the case of flat profiles. The precise value of
$\tan\beta$ depends on the size and shape of compact space. 

This paper is organized as follows. In section~\ref{sec:orbifoldGUT} we will discuss a class of simple
orbifold GUT models in which $y_t=g$ at tree level.
Section~\ref{sec:Corrections} is devoted to (quantum) corrections to this
relation. In section~\ref{sec:StringRealization} we will discuss explicit string theory realizations of
these settings. Phenomenological implications are described in
section~\ref{sec:Pheno}. Finally, section~\ref{sec:Conclusions} contains our
conclusions.

\section{Gauge--top unification in extra dimensions}
\label{sec:orbifoldGUT}

\begin{figure}[!h!]
\begin{center}
\CenterEps{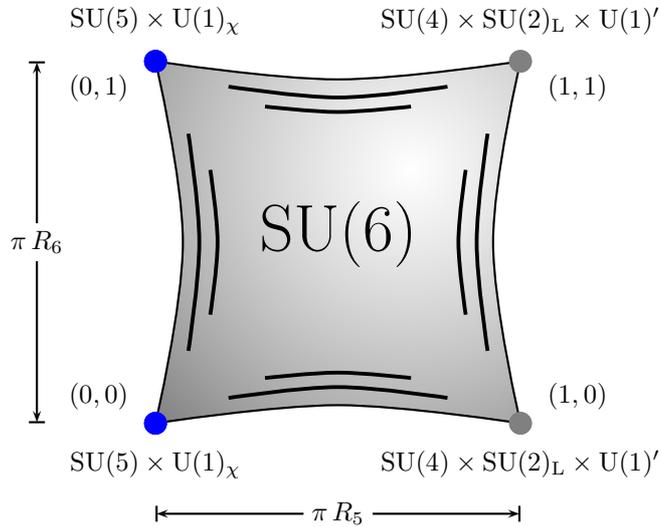}
\end{center}
\caption{6D orbifold GUT. We show the local gauge groups at the fixed points,
labeled by the localization quantum numbers $(n_2,n_2')$.}	
\label{fig:orbifoldGUT}	
\end{figure}

Let us now consider field-theoretic orbifold GUT settings in which $y_t$ arises
from gauge interactions. We shall focus on models with two extra dimensions and
an \SU6 bulk gauge group. We will also consider further \U1 factors, as their
presence has important implications for the prediction of gauge and Yukawa
couplings. The geometry of the model is $\mathbbm{T}^2/\mathbbm{Z}_2$. The
resulting orbifold can be envisaged as a ravioli or pillow, whose corners
correspond to the four fixed points. We label the fixed points by two integers
$n_2$ and $n_2'$ which take the values 0 or 1 (see
figure~\ref{fig:orbifoldGUT}). The lengths of the edges of the pillow or the
distances of the fixed points are given by $\pi\, R_5$ and $\pi\,R_6$,
respectively, where $R_5$ and $R_6$ denote the radii of the underlying 2-torus
$\mathbbm{T}^2$. We restrict ourselves to rectangular tori, leaving the general
case for future analysis.

The \SU6 bulk gauge group gets broken to $\SU5\times\U1_\chi$ and
$\SU4\times\SU2_\mathrm{L}\times\U1'$ at two inequivalent fixed points, i.e.\
fixed points with different $n_2$. The low-energy gauge group emerges as the 
intersection of these local gauge groups, and is given by
\begin{equation}
 G_\mathrm{SM}~=~\SU3_C\times\SU2_\mathrm{L}\times\U1_Y
\end{equation}
plus an additional \U1 factor. 
In orbifold GUT language, this is a consequence of the \Z2 boundary conditions
\begin{equation}\label{eq:BoundaryConditions}
 P~=~\diag(1,1,1,1,1,-1)\quad\text{and}\quad P'~=~\diag(1,1,1,-1,-1,1)\;,
\end{equation} 
which are to be imposed at the fixed points with $n_2=0$ and $n_2=1$,
respectively. In the description common to string theory, the difference between
the two local boundary conditions can be ascribed to the presence of a discrete
order 2 Wilson line in $y_5$ direction. The emerging symmetry breaking pattern
has been studied in the context of a 5D orbifold GUT~\cite{Burdman:2002se}, and
happens to arise in a 6D orbifold GUT limit of promising string
compactifications \cite{Buchmuller:2007qf,Ratz:2007my,Nilles:2008gq}. The bulk
supersymmetry in six dimensions corresponds to $N=2$ supersymmetry from a 4D
perspective. The 6D gauge multiplet $(V,\Phi)$ contains the 4D vector field $V$
as well as the chiral field $\Phi$. The boundary conditions
\eqref{eq:BoundaryConditions} result in orbifold parities for the components of
$\Phi$ that are opposite to those of $V$,
\begin{eqnarray}
 \lefteqn{\Phi ~=~\Phi^a\,\mathsf{T}_a\label{eq:DecompositionPhi}} \\
 & = &
 \footnotesize{\left(\begin{array}{ccc}
  \Phi_{(\boldsymbol{8},\boldsymbol{1})_0}^{(--)}
  -\frac{1}{\sqrt{15}}\Phi_Y^{(--)}
  +\frac{1}{2\sqrt{15}}\Phi_\chi^{(--)}
  & \frac{1}{\sqrt{2}}\Phi_{(\boldsymbol{3},\boldsymbol{2})_{-5/6}}^{(-+)}
  & \frac{1}{\sqrt{2}}\Phi_{(\boldsymbol{3},\boldsymbol{1})_{-1/3}}^{(+-)}
  \\
  \frac{1}{\sqrt{2}}\Phi_{(\overline{\boldsymbol{3}},\boldsymbol{2})_{5/6}}^{(-+)}
  & 
  \Phi_{(\boldsymbol{1},\boldsymbol{3})}^{(--)}
  +\frac{3}{2\sqrt{15}}\Phi_Y^{(--)}
  +\frac{1}{2\sqrt{15}}\Phi_\chi^{(--)}
  &
  \frac{1}{\sqrt{2}}\Phi_{(\boldsymbol{1},\boldsymbol{2})_{1/2}}^{(++)}
  \\
  \frac{1}{\sqrt{2}}\Phi_{(\overline{\boldsymbol{3}},\boldsymbol{1})_{1/3}}^{(+-)}
  &
  \frac{1}{\sqrt{2}}\Phi_{(\boldsymbol{1},\boldsymbol{2})_{-1/2}}^{(++)}
  & 
  \frac{-5}{2\sqrt{15}}\Phi_{(\boldsymbol{1},\boldsymbol{1})_{0}}^{(--)}
 \end{array}\right)\;.\nonumber}
\end{eqnarray}
Here we decompose the generators of \SU6 into the generators of
$G_\mathrm{SM}\times\U1_\chi$ and elements of the coset, using an obvious
notation.
At the massless level, only the doubly even states 
\begin{equation}
 h_u~=~\Phi_{(\boldsymbol{1},\boldsymbol{2})_{1/2}}^{(++)}
 \quad\text{and}\quad
 h_d~=~\Phi_{(\boldsymbol{1},\boldsymbol{2})_{-1/2}}^{(++)}
\end{equation}
are retained, which carry the quantum numbers of the MSSM Higgs doublets. The
normalization factors in \eqref{eq:DecompositionPhi} are a consequence of the
usual condition on the generators,
\begin{equation}
 \tr(\mathsf{T}_a\,\mathsf{T}_b)~=~\frac{1}{2}\delta_{ab}\;.
\end{equation}
Both $\Phi=\Phi^a\,\mathsf{T}_a$ and the various multiplets appearing on the
right-hand side of \eqref{eq:DecompositionPhi} are canonically normalized.

The settings further contain a bulk hypermultiplet $H$ that transforms as a 
$\boldsymbol{20}$-plet under \SU6. In $N=1$ language
\begin{equation}
 H~=~(\varphi,\varphi^c)\;,
\end{equation}
where $\varphi$ transforms as $\boldsymbol{20}$ (3-index antisymmetric tensor)
and $\varphi^c$ as $\boldsymbol{\overline{20}}$ under \SU6. After orbifold
projection, $\varphi$ gives rise to one copy of quark doublets ($q_3$) and
$\varphi^c$ leads to one superfield transforming as a $u$-type quark ($\bar
u_3$), as well as a lepton singlet $\bar e_3$. We refrain from specifying the
origin of the remaining SM matter here. Later, in
section~\ref{sec:StringRealization}, we will clarify this issue in the context
of string-derived models, where all anomaly constraints are automatically
fulfilled.

In order to extract the top Yukawa coupling it is sufficient to work with
component fields. We will then denote by $\xi$ and $\eta$ the two-component
spinors contained respectively in $\varphi$ and $\varphi^c$, and the scalar
component of $\Phi$ is given by
\begin{equation}
 \phi~=~\frac{A_5+\I A_6}{\sqrt{2}}\;.
\end{equation}
The 4D Yukawa term
\begin{equation}
 y_t\,\bar u_3\,q_3\,h_u~\subset~\mathscr{L}_\mathrm{4D}
\end{equation}
originates from the gauge interactions
\begin{equation}
 \eta\,(\,\overline{\partial}+\sqrt{2}g_6\,\phi)\,\xi
 ~\subset~\mathscr{L}_\mathrm{6D}
\end{equation}
with $g_6$ denoting the 6D gauge coupling and
$\overline{\partial}=\partial_5+\I\,\partial_6$ (cf.\
\cite{Arkani-Hamed:2001tb}).

In \SU{n} notation (see e.g.~\cite{Georgi:1975qb}), one has specifically for the
\SU6 case under consideration
\begin{eqnarray}
\lefteqn{
 \frac{1}{3!}\,\eta_{i_1i_2i_3}\left[
 \delta^{i_1}_{j_1}\,\delta^{i_2}_{j_2}\,\delta^{i_3}_{j_3}\,\overline{\partial}
 +\sqrt{2}\,g_6\,
 \left(
 	\phi^{i_1}_{j_1}\delta^{i_2}_{j_2}\,\delta^{i_3}_{j_3}
	+
	\delta^{i_1}_{j_1}\,\phi^{i_2}_{j_2}\,\delta^{i_3}_{j_3}
	+
	\delta^{i_1}_{j_1}\,\delta^{i_2}_{j_2}\,\phi^{i_3}_{j_3}
 \right)
 \right]\,\xi^{j_1j_2j_3}
 }\nonumber\\
 & = &
 \frac{1}{3!}\,\eta_{i_1i_2i_3}\,\left[
 	\delta^{i_3}_{j_3}\,\overline{\partial}
	+3\,\sqrt{2}\,g_6\,\phi^{i_3}_{j_3}
 \right]\,\xi^{i_1i_2i_3}\;.
\end{eqnarray}
$q_3$ is contained in the \SU5
$\boldsymbol{10}$-plet from the \SU6 $\boldsymbol{20}$-plet $\xi$,
\begin{equation}
 \boldsymbol{10}^{ij}~=~\xi^{ij6}\;,
\end{equation}
while $\bar u_3$ arises from the $\boldsymbol{10}$-plet in the \SU6
$\boldsymbol{\overline{20}}$-plet $\eta$,
\begin{equation}
 \frac{1}{2}\varepsilon_{ijk\ell m}(\boldsymbol{10}^c)^{\ell m}
 ~=~\eta_{ijk}\;.
\end{equation}
Here the indices $i,j,k\dots$ run from 1 to 5; the above implies that $q_3$ and
$\bar u_3$ do not stem from the same \SU5 $\boldsymbol{10}$-plet. 

We are now left with 
\begin{eqnarray}
 \sqrt{2}\,g_6\frac{1}{2}\eta_{ijk}\,\phi^k_6\,\xi^{ij6}
 & = & 
 \frac{\sqrt{2}\,g_6}{4}\varepsilon_{ijk\ell m}\,(\boldsymbol{10^c})^{ij}\,
 \boldsymbol{10}^{k\ell}\,\phi_6^m
 \nonumber\\
 & \supset & 
 \frac{g_6}{2}\varepsilon_{abc}\,\varepsilon_{\alpha\beta}\,\varepsilon^{abd}\,
 (\bar u_3)_d\,(q_3)^{c\alpha}\,(h_u)^\beta
 \nonumber\\
 & = & g_6\,\bar u_3\,q_3\,h_u\;.
\end{eqnarray} 
Here $a,b,c,d$ are $\SU3_C$ indices and $\alpha,\beta$ are $\SU2_\mathrm{L}$
indices. This calculation shows that $y_{t}=g$ at tree level, and confirms an
earlier computation~\cite{Buchmuller:2007qf}  (see also~\cite{Burdman:2002se}).

\section{Corrections to the equality of $\boldsymbol{y_t}$ and $\boldsymbol{g}$}
\label{sec:Corrections}

Having seen that $y_t=g$ at tree level, we now turn to discussing corrections to
this relation. Such corrections are of different origin. First, there are
radiative, logarithmic contributions to the wave function renormalization
constants coming from fields localized at the fixed points (or `branes') which
do not respect the unified bulk gauge symmetry. Second, the top Yukawa coupling
emerges from the $3\times3$ Yukawa matrix $Y_u$ by diagonalization. Subdominant
entries in $Y_u$ can therefore shift this eigenvalue. Third, there are
corrections coming from non-trivial localization properties of the bulk
hypermultiplets. As we shall see, this effect yields generically the numerically
dominant correction.

\subsection{Corrections from localized states}

The 6D $\beta$-function is comprised of a `power-law' and logarithmic piece,
\begin{equation}
 \beta_{6\text{D}}~=~b_6\,\mu^2\,R_5\,R_6+b_4\;.
\end{equation}
We have verified that the power-law pieces are universal for gauge and top
couplings, as it should be.
Assuming only the minimal matter content in the bulk to form the
third family of the MSSM, one obtains
\begin{equation}
 b_6^t~=~b_6^i~=~-4\quad\text{for}~1\le i\le3\;.
\end{equation}
The logarithmic corrections are sensitive to fields
sitting at the fixed points, where the \SU6 bulk symmetry is broken. The $b_4$
coefficients might hence not be universal. One might think of these corrections
as wave function renormalization constants localized at the fixed points. Their
impact can be estimated as
\begin{equation}
 \left|\Delta y_t-\Delta g_i\right|_\mathrm{log}
 ~\sim~\left|\Delta g_i-\Delta g_j\right|_\mathrm{log}
 ~\sim~\left|\Delta b_4
 \,\ln\left(\Lambda/M_\mathrm{GUT}\right)\right|\;,
\end{equation}
where $\Lambda$ denotes the cut-off and $\Delta b_4$ denotes the respective
difference of $\beta$-functions. These corrections are expected to be
numerically similarly relevant as MSSM threshold corrections, which originate
from the squarks and sleptons having masses that differ by an
$\mathcal{O}(1-10)$ factor. Such effects will not be studied in detail in the
present analysis.

\subsection{Diagonalization effects}

In general, we will expect the $u$-type Yukawa matrix to be of the form
\begin{equation}\label{eq:Yu-expected}
 Y_u~=~\left(\begin{array}{ccc}
  0 & 0 & 0\\ 0 & 0 & 0 \\ 0 & 0 & \mathcal{O}(g)
 \end{array}\right)
 +
 \left(\begin{array}{ccc}
  s^{n_{11}} & s^{n_{12}} & s^{n_{13}} \\
  s^{n_{21}} & s^{n_{22}} & s^{n_{23}} \\
  s^{n_{31}} & s^{n_{32}} & s^{n_{33}} 
 \end{array}\right)\;,
\end{equation}
where $s$ denotes the typical expectation value of some singlet field,
$n_{ij}\in\mathbbm{N}$, and we suppressed coefficients. It is clear that, after
a bi-unitary diagonalization, the top Yukawa coupling will be given by
\begin{equation}
 y_t~=~\mathcal{O}(g)+\mathcal{O}(s^n)
\end{equation}
with some $n\in\mathbbm{N}$. However, if the singlet scale and the exponents
$n_{ij}$ in \eqref{eq:Yu-expected} are such as to give rise to realistic Yukawa
couplings for the $u$ and $s$ quarks, this effect is expected to be negligible.
In fact, as we shall see next, localization effects in the compact dimensions
will generically have sizable impact on $y_t$, completely overwhelming the
effect described in the present subsection.

\subsection{Localization effects}
\label{sec:Localization}

\begin{figure} 
  \centering  
    \includegraphics[width=8cm]{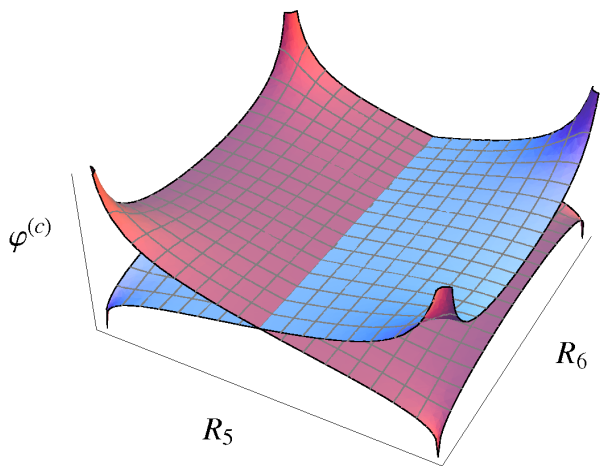}  
  \caption{Localization of two wavefunctions $\varphi$ and $\varphi^c$ 
           with opposite charges in the extra dimensions.
}
  \label{fig:overlap}
\end{figure}

Building on earlier analyses of the 5D
case~\cite{GrootNibbelink:2002qp,GrootNibbelink:2003gb}, Lee, Nilles and Zucker
(LNZ) \cite{Lee:2003mc} reported an interesting observation. Suppose there is a
\U1 symmetry, called $\U1_\mathrm{LNZ}$ in what follows, with localized
Fayet-Iliopoulos (FI) terms, i.e.\ 
\begin{equation}
 \tr(q_I)~\ne~0\;,
\end{equation}
where the trace extends over the states localized at the fixed points, labeled
by $I=(n_2,n_2')$. It turns out that then bulk states with non-zero
$\U1_\mathrm{LNZ}$ charge develop a non-trivial profile in the extra dimensions,
cf.\ figure~\ref{fig:overlap}.
In such a situation, the effective four-dimensional top Yukawa coupling, which
is given by the overlap of the corresponding wave functions, generically gets
reduced. It is amusing to see that the presence of fixed points, which
break $N=2$ supersymmetry down to $N=1$, also allows us to discriminate between
the $\varphi$ and $\varphi^c$ components of the hypermultiplet.

In \cite{Lee:2003mc} it was shown that the profile for the zero modes on
the 6D orbifold $\mathbbm{T}^2/\mathbbm{Z}_2$ is given by
\begin{equation}\label{eq:phi+}
 \psi
 ~\simeq~ 
 f\,\prod_{I}
 \left|\vartheta_1\left(\left.\frac{z-z_I}{2\pi}
 \right|\tau\right)\right|^{\frac{1}{2\pi}\,g_6\,q_\psi\,\xi_I}
 \exp\left(-\frac{1}{8\pi^2\tau_2}g_6\, q_\psi\, \xi_I\,(\im(z-z_I))^2 \right) \;,
\end{equation}
where we neglected a subleading logarithmic contribution, and $f$ is a
normalization factor. We explicitly checked
that this profile is the same for bosonic and fermionic degrees of freedom. The
complex coordinate $z$ is defined as
\begin{equation}
 z~=~\frac{1}{R_5}x^5+\frac{\tau}{R_6}x^6
\end{equation}
with $\tau$ denoting the modular parameter of the torus. In our settings we
assume the torus to be rectangular,  $\tau~=~\I\,\tau_2~=~\I\,R_6/R_5$. The
torus $\vartheta$-functions  can be written as~\cite{Mum:theta}
\begin{equation}
 \vartheta_1(z|\tau)
 ~=~
 \sum_{n\in\mathbbm{Z}}\mathrm{e}^{\I\,\pi\,\tau\,\left(n+\frac{1}{2}
 \right)^2+2\pi\,\I\,\left(n+\frac{1}{2}\right)\,\left(z+\frac{1}{2}\right)} \;.
\end{equation}
Furthermore, 
\begin{equation}\label{eq:xiI}
 \xi_I~ =~ \frac{1}{16\pi^2} g_6\, \Lambda^2\, 
 \left(\frac{1}{4}\tr(q)+\tr(q_I)\right)
\end{equation}
is the quadratically divergent piece of the FI term with $q$ the charges of the
bulk fields under the corresponding \U1, $q_I$ the charges of fields localized
at the fixed point $I$ and $\Lambda$ the cutoff of the theory. We will ignore
the first term in the sum of the right-hand side of \eqref{eq:xiI}, i.e.\ we
assume $\tr q=0$. We further take $\Lambda$ to be the higher-dimensional Planck
scale, $\Lambda^2 = M_\mathrm{P}/\sqrt{V_{56}}$ with $M_\mathrm{P}=2.43 \cdot
10^{18}\,\mathrm{GeV}$ and $V_{56}=2\pi^2R_5 R_6$ the volume of the extra
dimensions. In addition we have the usual relation between the
higher-dimensional gauge coupling and its four-dimensional counterpart,
$g_6=\sqrt{V_{56}}g$.  The constant $f$ can be determined through the
normalization condition
\begin{equation}
 \int_0^{\pi R_5}\!\D x_5\, \int_0^{2\pi R_6}\!\D x_6\, 
 |\psi|^2 ~=~ 1 \;.
\end{equation}

In our case we consider the orbifold  $\mathbbm{T}^2/\mathbbm{Z}_2$ with one
Wilson line in the $y_5$ direction. This means that there are two pairs of
equivalent fixed points, so all FI terms will have the same absolute value,
given that the effective FI term in four dimensions vanishes. Then the lightest
Kaluza-Klein masses are  either $M=1/(2R_{5})$ or $M=1/R_{6}$, due to the
presence of the Wilson line. Since we are interested in the limit $R_5 \ge R_6$
and since the GUT group gets broken by the Wilson line, we identify the GUT
scale with the corresponding Kaluza-Klein mass, $M_\text{GUT}=1/(2R_5)$. 
\begin{figure}[h]   
  \centering  
    \includegraphics[width=8cm]{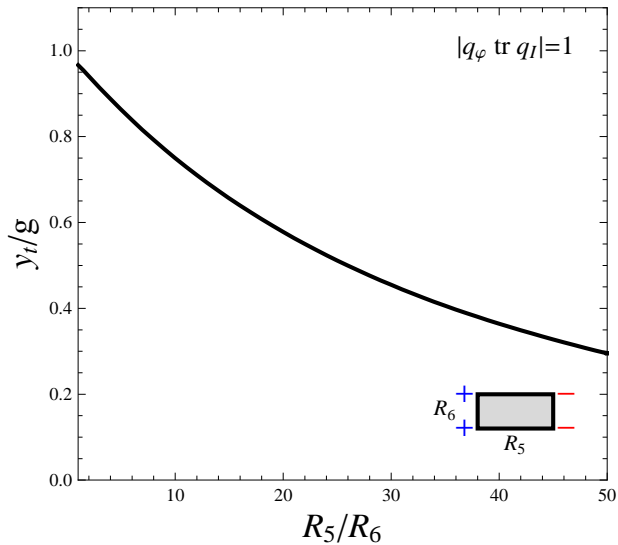}  
  \caption{Suppression of the top Yukawa coupling $y_t$ relative to the gauge
		   coupling $g$ due to the localization effects as a function of the
		   anisotropy.  The inlay indicates that the region $R_5>R_6$,
		   where the distance between equivalent fixed points with coinciding
		   traces is smaller that the one between inequivalent fixed points, is
		   considered. In the other limit, i.e.\ for $R_6\gg R_5$, there is no analogous suppression.
}
  \label{fig:reduction}
\end{figure}

Let us now briefly compare the terms giving rise to the top Yukawa and gauge
couplings, respectively. While the first is proportional to the overlap integral
over $h_u\,q_3\, \bar{u}_3$, the latter scales like the integral over $A\, \bar
u_3\,\bar u_3^\dagger$. The Higgs field itself retains a flat profile since it
comes from a non-Abelian gauge multiplet.
Taking further into account that the profile of $\bar u_3^{\dagger}$ is
identical to the one of $\bar u_3$, while the profiles of $q_3$ and $\bar u_3$
are inverse to each other, it is clear that the top Yukawa coupling will always
be reduced with respect to the gauge coupling, when the correct normalization is
taken into account. This effect is illustrated in figure~\ref{fig:reduction}. 
As can be seen, the reduction depends on the given charges and is more
pronounced for  anisotropic compactifications. This behavior is to be expected,
since in the limit $R_6 \rightarrow 0$ the 5D case should be recovered, where
the bulk fields effectively become brane fields and hence the overlap should
vanish. This is because in this limit the bulk fields get exponentially
localized towards the opposite ends of the
interval~\cite{GrootNibbelink:2002qp}, such that the overlap becomes
exponentially suppressed.

\section{Explicit string theory realization}
\label{sec:StringRealization}

Our analysis is motivated by recent progress in string model building
\cite{Buchmuller:2005jr,Buchmuller:2006ik,Lebedev:2006kn,Buchmuller:2007qf,Lebedev:2007hv,Buchmuller:2008uq},
where $\mathcal{O}(100)$ explicit (and globally consistent) models with the
exact spectra of the MSSM, the so-called `heterotic mini-landscape', have been
derived. In a subclass of these models  the top (but neither the bottom nor the
$\tau$) Yukawa coupling is related to the gauge coupling. This applies in
particular to the two models that have been analyzed in some detail,
\cite{Buchmuller:2005jr,Buchmuller:2006ik,Buchmuller:2008uq,Buchmuller:2007qf}
and the `benchmark model 1A' in \cite{Lebedev:2007hv}. 
In what follows, we will study gauge-top unification in these models.
Our analysis should be viewed as the first step towards a full string theory
calculation, where one computes the FI term in string theory, takes into account
arbitrary values for the torus parameter $\tau$ and the full volume of the six
compact dimensions. In the present study, we will restrict ourselves to the
somewhat naive orbifold GUT picture, and will take the cutoff $\Lambda$ to be
the 6D Planck scale, as before.

\subsection{Motivation of the orbifold GUT picture}

Taking the orbifold GUT limit of a string compactification has a rather profound
motivation. Witten proposed in a footnote \cite[footnote~3]{Witten:1996mz} a
possible way to explain the discrepancy between the string and the GUT scales.
This can be accomplished by considering highly anisotropic
compactifications, where $M_\mathrm{GUT}$ is associated to the inverse of the
largest radius, while all (or most of) the other radii are much smaller. In this
case, the volume of compact space can be small enough to ensure that the
perturbative description of the setting is still appropriate. This idea has been
studied in some detail more recently \cite{Hebecker:2004ce}. The outcome of the
analysis is that the above puzzle can be resolved if the largest radius
is by a factor 50 or so larger than the other radii. The question of how to
stabilize the largest radius has also been addressed in the framework of field
theory, and it has been found that one can indeed obtain
$R_5\simeq1/(2M_\mathrm{GUT})$~\cite{Buchmuller:2008cf,Gross:2008he,Buchmuller:2009er}.
Also here localized FI terms can play an important role.
The question why one radius behaves so differently from the others will be
addressed elsewhere~\cite{Kappl:2009pr}. In what follows, we will build on these
results, and consider 6D orbifold GUT limits with particular emphasis on the
highly anisotropic case $R_5\gg R_6$. The transverse four dimensions will not be
considered.

\subsection{A specific example}
\label{sec:BenchmarkModel}

\begin{figure}[h]
\centerline{\includegraphics{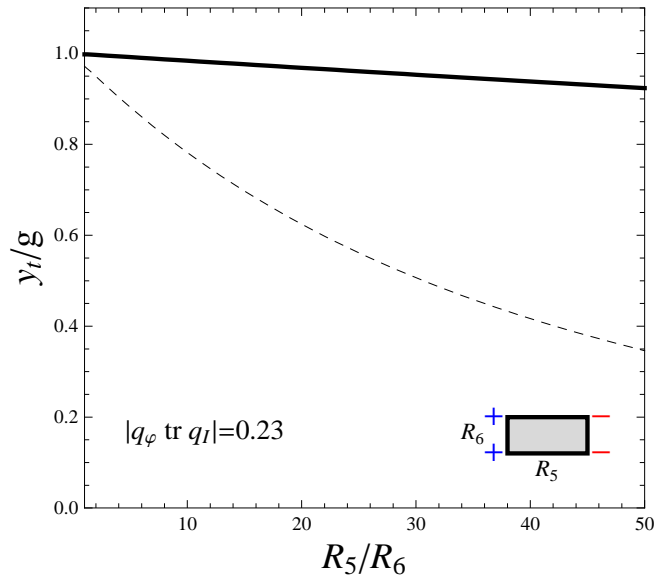}}
\caption{Estimate of $y_t/g$ in the benchmark model 1A of \cite{Lebedev:2007hv}.
For comparison, the dashed line shows the suppression which occurs with an
increased cut-off, $\Lambda \rightarrow 2 \Lambda$. }
\label{fig:PlotBenchmarkModel}
\end{figure}

For concreteness, let us first focus on the benchmark model 1A of
\cite{Lebedev:2007hv}. 
In order to be in accord with the discussion in
section~\ref{sec:Localization} (and the analysis by LNZ~\cite{Lee:2003mc}),
here and below we take only \U1 factors orthogonal to the
so-called `anomalous \U1' direction $\mathsf{t}_\mathrm{anom}$ into account.
The general case will be explored elsewhere. 
One can choose the basis of these \U1s such that the local FI terms at the fixed
points are entirely in one \U1 direction, called $\U1_\mathrm{LNZ}$ in
what follows. The corresponding generator reads
\begin{equation}
 \mathsf{t}_\mathrm{LNZ}~=~
 \frac{1}{\sqrt{105}}
  \left(-\tfrac{3}{2} , 0 , 0 , 0 , 0 , 0 , 0 , 0\right)\, 
 \left(-\tfrac{11}{4} , \tfrac{21}{4} , 0 , \tfrac{11}{4} , 0 , \tfrac{11}{4} , 0 , 0
 \right)\;,
\end{equation}
and is normalized such that $|\mathsf{t}_\mathrm{LNZ}|^2=1/2$. In this
normalization, the charges of $\varphi$ and $\varphi^c$, containing $q_3$ and
$\bar u_3$, are
\begin{equation}
 q_\varphi~=~-q_{\varphi^c}
 ~=~\frac{1}{\sqrt{105}}\frac{3}{4}\;,
\end{equation}
respectively. One further has
\begin{equation}\label{eq:LocalizedTracesBenchmarkModel}
 \tr_{\SU5~\mathrm{brane}} q_\mathrm{LNZ}~=~\frac{64}{\sqrt{105}}
 \quad\text{and}\quad
 \tr_{\SU4\times\SU2~\mathrm{brane}} q_\mathrm{LNZ}~=~-\frac{64}{\sqrt{105}}\;,
\end{equation}
where the \SU5 and $\SU4\times\SU2$ branes comprise the fixed points with
$n_2=0$ and 1, respectively, 
with the fixed points carrying half of the trace. We can hence apply the
results of section~\ref{sec:Localization} in order to evaluate $y_t$.  The
ratio $y_t/g$ for the present example is shown in
figure~\ref{fig:PlotBenchmarkModel}. For this figure, we used our field
theoretic estimate of the FI term, i.e.\ chose to identify $\Lambda$ with the 6D
Planck scale, as before.
As discussed in section~\ref{sec:Localization}, the localization effect becomes
more pronounced when the ratio $R_5/R_6$ increases. We also show what the
suppression would be if we chose an increased cut-off,
$\Lambda^2=4M_\mathrm{P}/\sqrt{V_{56}}$, designed in such a way that for
$R_5=50\,R_{\ge 5}$ $\Lambda$ equals the heterotic string scale, 
$M_\mathrm{string} \simeq 8 \cdot10^{17}\,\mathrm{GeV}$, as in the proposal
discussed by Hebecker and Trapletti~\cite{Hebecker:2004ce}.

It is amazing to see that the top Yukawa coupling gets only reduced if we take
the `more appealing' orbifold GUT limit $R_5\gg R_6$. If we were to choose $R_6\gg
R_5$, which would lead us to an orbifold GUT with $\SU3\times\SU3$ bulk group
that gets broken to the SM at the two equivalent boundaries, the top Yukawa
coupling would no longer be suppressed since $\tr q_I$ vanishes at both
boundaries.

\subsection{Mini-landscape survey}

We have repeated the analysis of section~\ref{sec:BenchmarkModel} for a subclass
of the heterotic mini-landscape models~\cite{Lebedev:2006kn,Lebedev:2007hv}, in
which
\begin{itemize}
 \item there is an orbifold GUT limit as described in
 section~\ref{sec:orbifoldGUT};
 \item it has been explicitly verified that exotics decouple consistently with
 supersymmetry, i.e.\ with vanishing $F$- and $D$-terms.
\end{itemize}
All these models, including the model derived in~\cite{Buchmuller:2005jr,Buchmuller:2006ik,Buchmuller:2008uq,Buchmuller:2007qf}
and the `benchmark model 1A' discussed before, are based on the same shift vector. 
They turn out to have the following family structure (up to
vector-like states):
\begin{itemize}
 \item $1^\mathrm{st}$ and $2^\mathrm{nd}$ families come from 
 $\boldsymbol{16}$-plets localized at \SO{10} fixed points, which correspond to
 the fixed points with $n_2=0$ in the orbifold GUT limit
 (figure~\ref{fig:orbifoldGUT});
 \item $3^\mathrm{rd}$ family $\bar d$ and $\ell$ (i.e.\ the $3^\mathrm{rd}$
 family $\overline{\boldsymbol{5}}$ in \SU5 language) come from the $T_{2/4}$
 twisted sectors and therefore are localized on two-dimensional submanifolds
 in compact 6D space;
 \item $3^\mathrm{rd}$ family $\bar u$, $\bar e$ and $q$ as well as the Higgs fields $h_u$
 and $h_d$ are bulk fields, i.e.\ free to propagate everywhere in compact space.
\end{itemize}
Only 4 out of 56 candidate models do not have localized FI terms.

Yukawa couplings connecting the Higgs fields to matter may be written as
overlap integrals. One could then expect that the couplings of the first two
generations are suppressed by the total 6D volume while the $\tau$ and $b$
Yukawas, $y_\tau$ and $y_b$, are suppressed by the volume of the 4D space
transverse to the two-dimensional submanifold, while the top Yukawa $y_t$ is
unsuppressed, as discussed before. This leads us to the hierarchy
\[
 \text{Yukawa couplings of the first two generations}
 ~\ll~
 y_\tau,y_b
 ~\ll~ y_t\;.
\]
It should be mentioned that these localization properties are highly non-trivial
(and come out `for free'). In fact, the cancellation of all 4D and
higher-dimensional gauge anomalies, which is guaranteed by modular
invariance~\cite{Vafa:1986wx}, appears rather miraculous in field theory
\cite{Buchmuller:2007qf}. Also discrete anomalies have been shown to
cancel~\cite{Araki:2008ek}. 

\begin{figure}[h]
 \centerline{\subfigure[$|q_\varphi\,\tr q_I|$ at $I=(0,0)$.]{\includegraphics[scale=0.6]{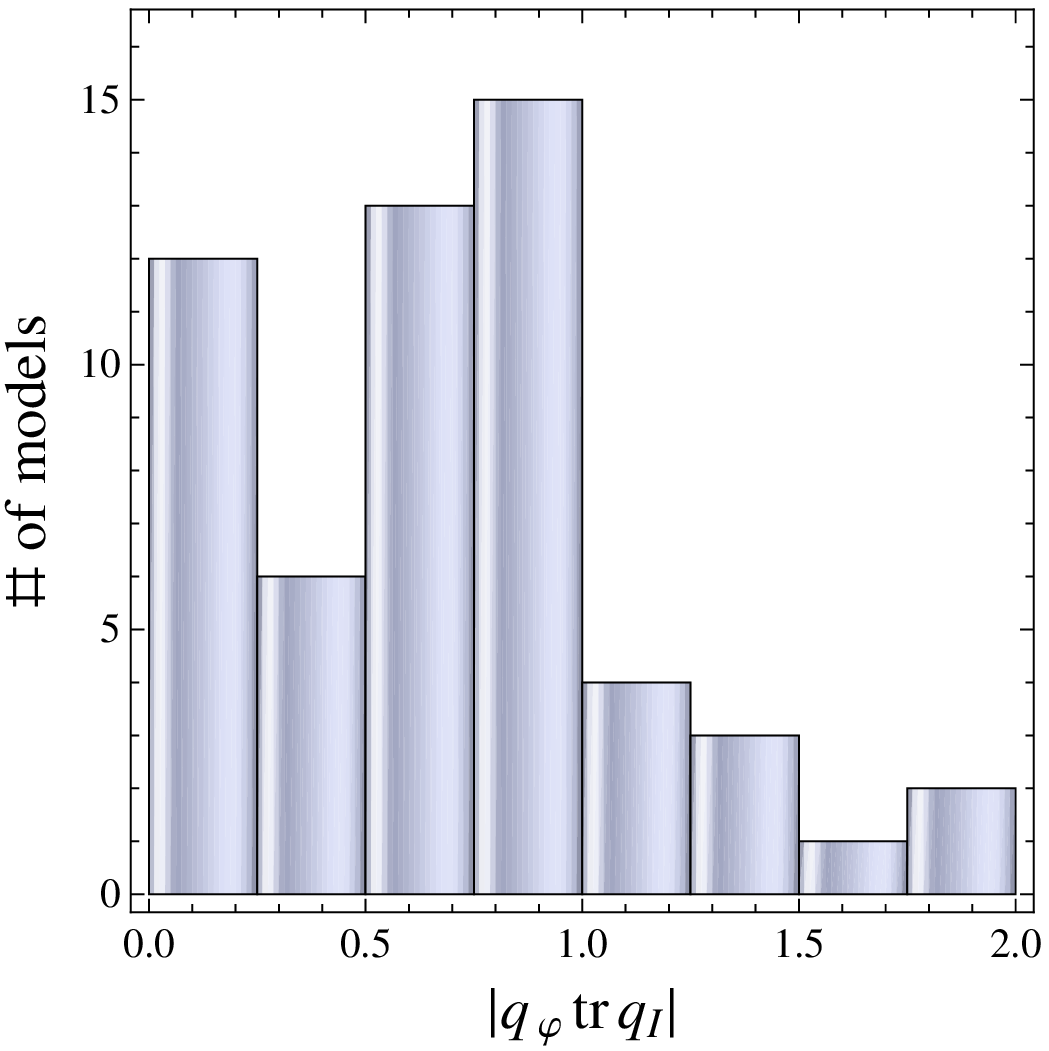}}\quad
 \subfigure[$y_t/g$.]{\includegraphics[scale=0.8]{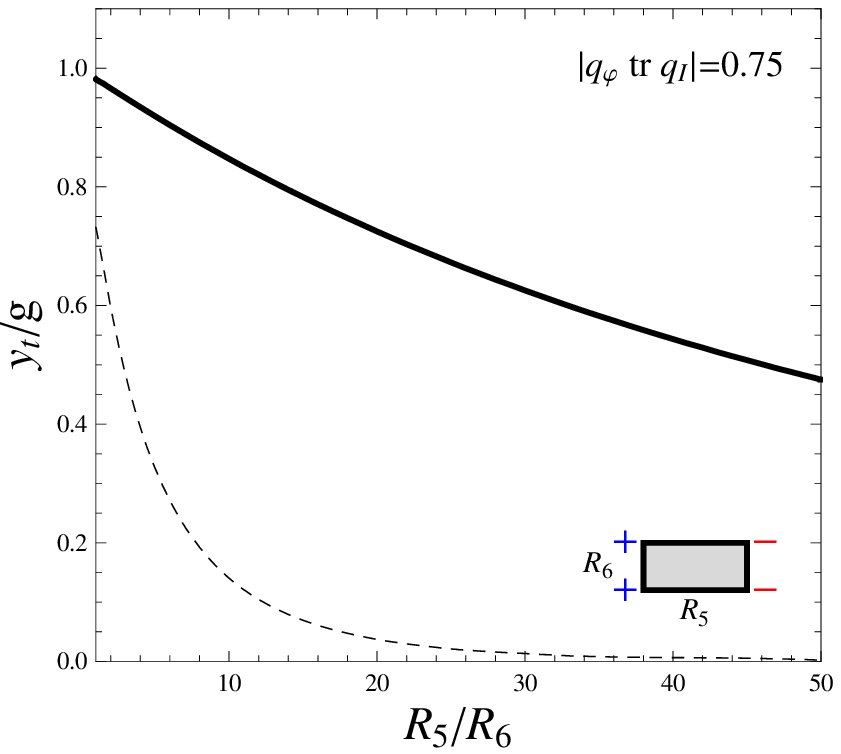}}}
\caption{Statistics of $|q_\varphi\,\tr q_\mathrm{LNZ}|_{n_2=n_2'=0}|$ in a
class of string-derived orbifold GUTs and suppression of the top Yukawa coupling for the
average value.}
\label{fig:statistics}
\end{figure}

Let us now come back to the question of the value of $y_t$ at the
compactification or GUT scale. As before in section~\ref{sec:BenchmarkModel}, it
is possible to rotate the basis of \U1 factors (orthogonal to
$\mathsf{t}_\mathrm{anom}$) such that there is precisely one \U1, which we again
call $\U1_\mathrm{LNZ}$, with the following properties:
\begin{itemize}
 \item the chiral multiplets $\varphi$ and $\varphi^c$ 
  have non-zero (and opposite) $\U1_\mathrm{LNZ}$ charges;
 \item  the traces
 $\tr_{\SU5~\mathrm{brane}} q_\mathrm{LNZ}$ and
 $\tr_{\SU4\times\SU2~\mathrm{brane}} q_\mathrm{LNZ}$
 have opposite sign (as in \eqref{eq:LocalizedTracesBenchmarkModel});
 \item $|\mathsf{t}_\mathrm{LNZ}|^2=\frac{1}{2}$, i.e.\ the charges are in ``GUT
 normalization''.
\end{itemize}
The statistics of the localized traces and the charges of the bulk
hypermultiplets $\varphi$ and $\varphi^c$ are depicted in
figure~\ref{fig:statistics}.

\section{Phenomenological implications}
\label{sec:Pheno}

Let us now
come to some phenomenological implications of the gauge-top unification scenario.  
Given a certain pattern of soft terms as well as the top Yukawa coupling at the GUT scale, 
the value of $\tan\beta$ can be determined.  Note that the value of $\tan\beta$ is also quite sensitive to the
mass of the top quark, where the latest experimental value is given by $m_t =
173.1 \pm 1.3\,\mathrm{GeV}$~\cite{:2009ec}. 

\begin{figure} 
\centerline{\subfigure[$y_t$ vs.\ $\tan\beta$.]{%
 \includegraphics[width=7cm]{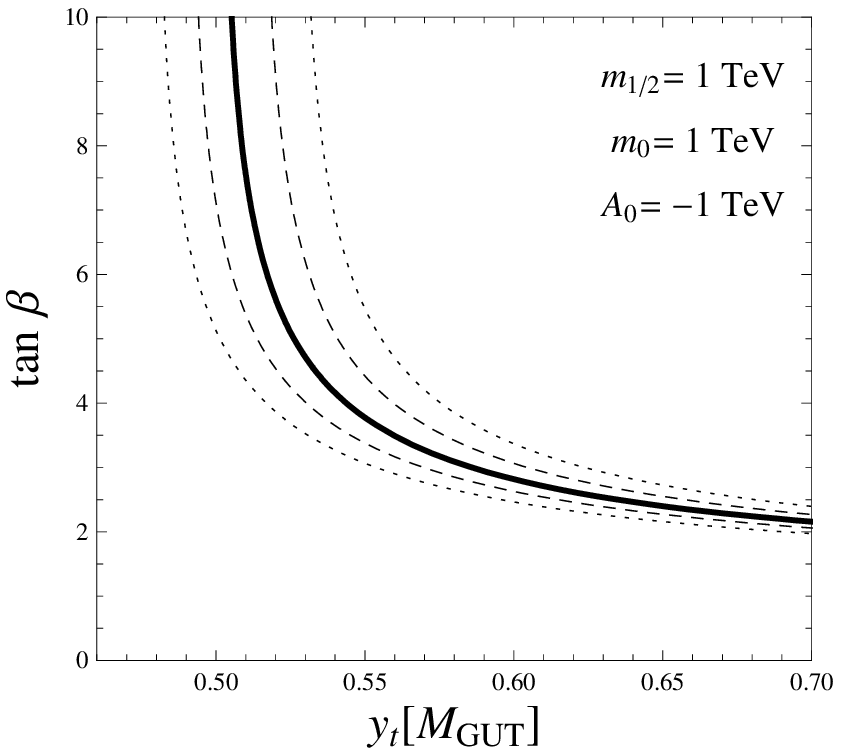}  
  }
\subfigure[$R_5/R_6$ vs.\ $|q_\varphi\,\tr q_I|$.]{%
    \includegraphics[width=7cm]{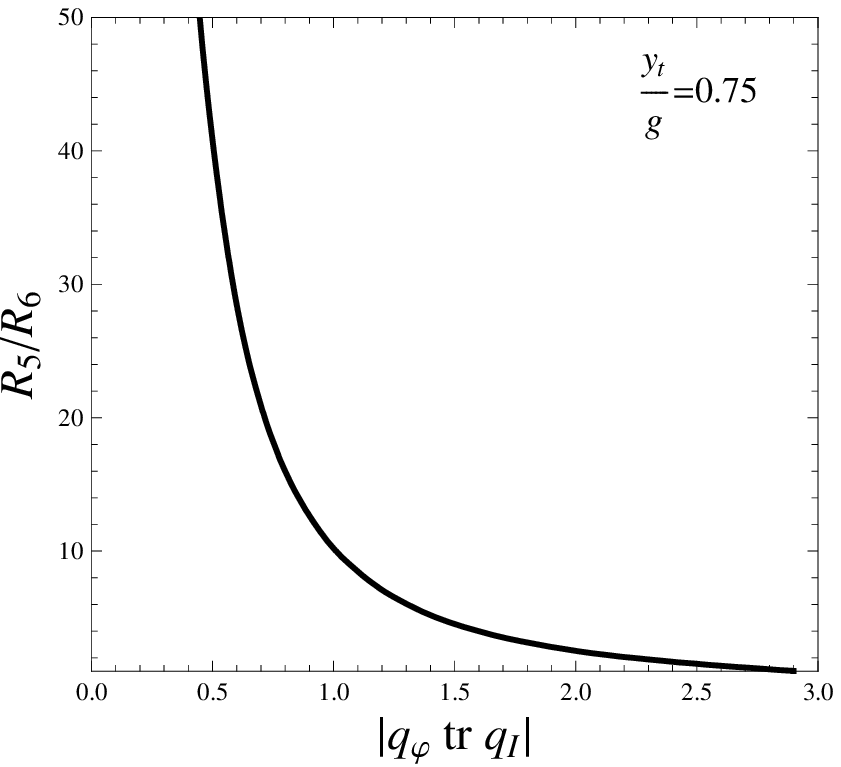}  
	}
}	
  \caption{(a) $\tan\beta$ as a function of the top Yukawa coupling at the GUT
		   scale. The central line corresponds to the central value of the top
		   quark mass, whereas the dashed and dotted lines correspond to the one
		   and two sigma intervals, respectively. A larger top mass results in a
		   larger $\tan\beta$.  
		   (b) shows for given $q_\varphi\, \tr q_I$ the anisotropy that is
		   needed to reduce $y_t/g$ to 0.75.
}
  \label{fig:ytvstanb}
\end{figure}

It turns out that the resulting value for $\tan\beta$ mainly depends on
$y_t(M_\text{GUT})$. To illustrate this dependence we plot the top Yukawa
coupling at the GUT scale versus $\tan\beta$ for different values of the top
quark mass in figure~\ref{fig:ytvstanb}. We impose a specific set of
boundary conditions at the GUT scale, the so-called mSUGRA set with
$m_0=m_{1/2}=-A_0=1\,\mathrm{TeV}$, and use SOFTSUSY~\cite{Allanach:2001kg} for
our numerical analysis.

The value of the unified gauge coupling on the other hand  turns out to be only
very weakly dependent on the soft parameters and is always close to $g \simeq
0.7$. This implies that, given the tree-level relation $y_t=g\simeq0.7$, the
resulting value for $\tan\beta$ is quite  small, $\tan\beta\simeq 2$. An
immediate question is whether such small values for $\tan\beta$ and hence the
tree-level relation are still valid, since the small $\tan\beta$ region is
probed experimentally by searches for the light Higgs boson
\cite{Heinemeyer:1999zf}. This tension can be seen by comparing the LEP bound on
the SM Higgs mass, $m_h \ge 114.4\,\mathrm{GeV}$, to the theoretical upper bound
on the lightest Higgs mass as a function of $\tan\beta$. At tree-level the
maximal mass for the lightest Higgs is given by $m_h^2 \simeq M_Z^2 \cos^2 2
\beta$, which vanishes for $\tan\beta=1$. Radiative corrections can
significantly increase the Higgs mass compared to the tree-level value, but
still $m_h$ is minimized for $\tan\beta$ around one in the MSSM. While the LEP
Higgs bound does not apply for all of the MSSM parameter space, in the small
$\tan\beta$ region it is applicable to a good approximation, since here the
lightest  CP-even Higgs boson couples to the $Z$ with SM-like strength
\cite{Degrassi:2002fi}. For the so called `$m_h^\text{max}$ scenario' within the
MSSM, designed such that for fixed values of $m_t$ and $M_\text{SUSY}$ the
predicted value of the lightest Higgs is maximized for each value of $\tan\beta$
and $m_A$, the lower bound on $\tan\beta$ is around 2 \cite{Degrassi:2002fi}.
Although hence values of $\tan\beta$  around 2 are still possible, one should
not forget that this  is possible only when the parameters are tuned
accordingly. In a large part of the parameter space $\tan\beta$ has to be
larger. To see what is natural as the `smallest value of $\tan\beta$ without too
much tuning', we restrict ourselves to mSUGRA. Taking
$m_{1/2}=m_0=-A_0=1\,\mathrm{TeV}$ and the top mass at its two sigma upper
bound, the predicted Higgs mass is above the LEP bound for $\tan\beta \gtrsim
3.3$.   This in turn implies a `natural range' for the top Yukawa coupling at
the GUT scale of $0.48\lesssim y_t \lesssim 0.6$, which translates into  $0.69
\lesssim y_t/g \lesssim 0.86$.  Note that in the presence of vector-like matter
(in complete \SU5 representations) with masses below the GUT scale the relative
suppression of $y_t$ has to be stronger.

In conclusion we can say that the naive picture of gauge-top unification, although not excluded, seems
to be possible for very special patterns of the soft terms only. 
However, we have seen in the previous sections that the top Yukawa coupling gets somewhat reduced 
with respect to the tree-level relation $y_t=g$, depending on the geometry of the extra-dimensional
space. Turning this around means that, given a value for $\tan\beta$, we gain access to the geometry
of the extra-dimensional space. In particular it seems that highly anisotropic compactifications are favored,
in accord with \cite{Hebecker:2004ce}.

Our findings seem to give another motivation for highly anisotropic string
compactifications. The top Yukawa coupling seems to give a preference
to this limit.

\section{Discussion}
\label{sec:Conclusions}

We have discussed orbifold GUT scenarios in which the top Yukawa coupling arises
from gauge interactions. This leads to the tree-level relation $y_t=g$ at the
GUT or compactification scale. It turns out that in scenarios with localized FI
terms $y_t$ is suppressed against $g$, which seems also to be required by data.
The suppression depends on the size of the localized FI terms, the charge of the
bulk hypermultiplets and, in particular, on the geometry of compact space. We
have analyzed string-derived models which reproduce the MSSM below the
compactification scale and which possess orbifold GUT limits with the above
features. This allowed us to get a feeling for what the FI terms and charges
`should be'. Using this as an input, we find that the observed value of the top
Yukawa coupling seems to favor anisotropic geometries, which might also be a key
ingredient for reconciling the scale of grand unification with the Planck or
string scales. This result might be interpreted as further support for the idea
of orbifold GUTs and, in particular, for their embedding in the heterotic
string.

As mentioned, our analysis should be viewed as a step towards getting
sufficiently accurate predictions from string theory. The precision of the
present analysis is mainly limited by our ignorance of the cut-off $\Lambda$,
for which we simply used the 6D Planck scale.  One can improve the accuracy
significantly by directly computing the FI terms in string theory. Another important question is
why, from a top-down perspective, one radius behaves so differently from the
others. We plan to clarify these issues in the near future.

\subsection*{Acknowledgments}
We would like to thank H.P.~Nilles, P.K.S.~Vaudrevange and M.~Winkler for useful
discussions and H.M.~Lee for correspondence. This research was supported by the
DFG cluster of excellence Origin and Structure of the Universe, and the
\mbox{SFB-Transregio} 27 "Neutrinos and Beyond" by Deutsche
Forschungsgemeinschaft (DFG). The work of P.H.\ was supported by the Alexander
von Humboldt foundation.

\bibliography{Orbifold}
\bibliographystyle{ArXiv}

\end{document}